\documentclass[letterpaper, conference]{IEEEtran}
\IEEEoverridecommandlockouts          

\usepackage[utf8]{inputenc}
\usepackage[T1]{fontenc} 
\usepackage{amsmath,amssymb,amsfonts,amsthm}
\usepackage{algorithmic}
\usepackage[ruled,vlined,linesnumbered]{algorithm2e}
\usepackage{graphicx}
\usepackage{textcomp}
\usepackage{xcolor}
\usepackage{float}
\usepackage{tabularx}
\usepackage{caption}
\usepackage{subcaption}
\usepackage{tikz}
\usepackage{epstopdf}
\usepackage{mathabx}
\usepackage{diagbox}
\usepackage{flushend}
\usepackage{stfloats}
\usepackage[letterpaper, left=0.68in, right=0.68in, bottom=1.0in, top=0.7in]{geometry}
\usetikzlibrary{patterns}
\usepackage{comment}
\def\BibTeX{{\rm B\kern-.05em{\sc i\kern-.025em b}\kern-.08em
    T\kern-.1667em\lower.7ex\hbox{E}\kern-.125emX}}

\begin{document}
\newcommand{\mluScip}[0]{\texttt{MLU-OPT}}
\newcommand{\mluLs}[0]{\texttt{MLU-LS}}
\newcommand{\mluFW}[0]{\texttt{MLU-FW}}

\newcommand{\jeremie}[1]{\textcolor{black}{#1}} 
\newcommand{\quang}[1]{\textcolor{black}{#1}} 

\newcommand{\allTNsScip}[0]{\texttt{ATNs-OPT}}
\newcommand{\allTNsLS}[0]{\texttt{ATNs-LS}}
\newcommand{\allTNsFW}[0]{\texttt{ATNs-FW}}
\newcommand{\allTns}[0]{\texttt{ATNs}}
\newcommand{\mlu}[0]{\texttt{MLU}}
\newcommand{\Dist}[0]{\texttt{Dist}}
\newcommand{\delayfunc}[3]{\texttt{Delay}^{#1}_{#2}\left({#3}\right)}
\newcommand{\lossfunc}[3]{\texttt{Loss}^{#1}_{#2}\left({#3}\right)}
\newcolumntype{P}[1]{>{\centering\arraybackslash}p{#1}}
\newcommand{\eval}[0]{\textit{eval()}}
\newcommand{\cursol}[0]{\texttt{CurSol}}
\newcommand{\curobj}[0]{\texttt{CurObj}}
\newcommand{\newobj}[0]{\texttt{NewObj}}

\title{\LARGE \bf
Global QoS Policy Optimization in SD-WAN
}

\author{Pham Tran Anh Quang$^{1}$, J\'er\'emie Leguay$^{1}$, Xu Gong$^{2}$, Xu Huiying$^{2}$\\
Huawei Technologies Ltd., $^{1}$Paris Research Center, France, $^{2}$ Dongguan Research Center, China. 
 }
\maketitle

\begin{abstract}  In modern SD-WAN networks, a global controller is able to steer traffic on different paths based on application requirements and global intents. However, existing solutions cannot dynamically tune the way bandwidth is shared between flows inside each overlay link, in particular when the available capacity is uncertain due to cross traffic. 
In this context, we propose a global QoS (Quality of Service) policy optimization model that dynamically adjusts rate limits of applications based on their requirements to follow the evolution of network conditions. It relies on a novel cross-traffic estimator for the available bandwidth of overlay links that only exploits already available measurements. We propose two local search algorithms, one centralized and one distributed, that leverage cross-traffic estimation. We show in packet-level simulations a significant performance improvement in terms of SLA (Service Level Agreement) satisfaction. For instance, the adaptive tuning of load balancing and QoS policies based on cross-traffic estimation can improve SLA satisfaction by $40\%$ compared to static policies.  
\end{abstract}

\section{Introduction}

Many enterprises are adopting Software-Defined Wide Area Network (SD-WAN)~\cite{yang2019software}  technologies to reduce costs and optimize performance. Most of them are migrating from fully dedicated private lines, using MSTP (Multi-service SDH-based Transport Platform), to a cost-efficient mix of MPLS-VPN (MV) and  Broadband Internet services. 
While the use of cheaper network services can yield to weaker Service Level Agreements (SLA) guarantees, high performance can still be obtained when combining them. In this context, to reach the best user experience, advanced traffic engineering mechanisms based on dynamic load balancing and adaptive queuing need to be designed. Indeed, the main challenge is to intelligently make use of all the available capacity to fulfill application requirements and optimize user experience.

In typical SD-WAN architectures, a centralized controller maintains a set of policies deployed at edge routers interconnecting multiple sites (e.g., enterprise branches, data centers). Each edge router is configured to steer traffic towards its peers over several transport networks. These routers are responsible for the load balancing of flows so as to meet application SLAs. The centralized controller can  optimize or influence policies to improve SLA satisfaction and global objectives, sometimes called
"intents"~\cite{pang2020survey}, that can be the minimization of financial expenses, of the congestion or the maximization of the experienced network quality.

Several solutions~\cite{yang2019software} have been proposed for the dynamic selection of paths in order  to satisfy SLA requirements. The general idea is to compare the quality of paths with application requirements and update the path selection strategy inside routers when needed. 
In a previous work~\cite{quangintent}, we introduced a routing policy optimization algorithm for Smart Policy Routing (SPR)~\cite{huawei-SPR} to optimize global intents (e.g., minimum cost, minimum latency) and meet QoS requirements. Despite that it helps to load balance traffic across multiple transport networks, bandwidth still needs to be efficiently shared between applications inside each of them.

To address this challenge, a number of adaptive queuing and Active Queue Management (AQM) techniques~\cite{aqmsurvey} have been proposed to help to meet SLA requirements in terms of delay and throughput. In particular, the dynamic adaptation of queue parameters, such as the weights in Adaptive Weighted Fair Queuing (AWFQ)~\cite{FRANTTI200911390, sayenko2006comparison}, has been shown to significantly improve  performance.
Nonetheless, existing mechanisms are local and work at the level of individual routers, without trying to globally improve  QoS. In~\cite{HUSSAIN20031143}, for instance, an agent at the destination informs  source nodes of delay violations, so that upstream agents adjust their queuing weights.
In this work, on the contrary, we propose to globally optimize queuing policies among edge routers sharing resources (e.g., bottleneck links).

Furthermore, one of the key issue in SD-WAN comes from the fact that traffic is only controlled from the edge and competing with cross-traffic inside each transport network. Therefore, to optimize policies, the use of  available bandwidth (ABW)~\cite{chaudhari2015survey} estimators becomes instrumental. 
Indeed, they can be used to learn about the available capacity and optimize policies accordingly. Most of the proposed methods need to inject
dedicated probing packets into the network and require an
additional software deployed at edge routers. To mitigate this, we propose a new estimator that only exploits already available measurements about latency and packet loss.

In this paper, we present a control plane architecture where edge devices decide about path selection and bandwidth sharing for the different applications based on policies decided by a centralized controller. On top of the optimization of SPR policies~\cite{quangintent} by the controller for load balancing, QoS policies can also be dynamically optimized, instead of being static. To periodically tune QoS policies, we present a centralized policy optimization algorithm that leverages on SABE (Silent Available Bandwidth Estimation), a new ABW estimator that only exploits already available measurements in terms of delay and packet loss, instead of injecting probing packets. We formulate a global QoS policy optimization problem that aims at protecting high priority flows and providing low priority ones with a fair sharing of the remaining bandwidth in case of heavy congestion. In order to optimize policies at large scale and in an incremental manner, our solution is based on a local search algorithm. \jeremie{We analyze the accuracy of SABE and evaluate its impact in packet level simulations on policy optimization. While in the tested scenario the relative error for the estimation of the available bandwidth is 10\% in average (and up to 20\%), we show that it can really boost performance in terms of SLA satisfaction.}
In addition, we design a distributed algorithm and show that the estimation of the available bandwidth can be used as an implicit signal for the collaboration between edge routers. In simulations, we demonstrate that the centralized solution achieves near-optimal performance in terms of SLA satisfaction, while the distributed one is close. Overall, the adaptive tuning \jeremie{and global optimization} of both SPR and QoS policies based on cross-traffic estimation can improve SLAs satisfaction rate by $40\%$ compared to static policies.

The rest of this paper is structured as follows. Sec.~\ref{sec:RElatedWork} reviews the related work. The system architecture and the problem formulation are introduced in Sec.~\ref{sec:SystemDescription} and~\ref{sec:ProblemFormulation}. 
Sec.~\ref{sec:ABW} provides a new available bandwidth estimation method.
Sec.~\ref{sec:Algorithms} introduces algorithms to optimize QoS policies. Sec.~\ref{sec:NumericalResults} presents our numerical results and Sec.~\ref{sec:conclusion} concludes this paper.

\section{Related Work}
\label{sec:RElatedWork}

Mechanisms for intelligent path selection have already been proposed for different purposes. Variants of ECMP (Equal-cost multipath)~\cite{Alizadeh:Conga, kabbani2014flowbender, he2015presto} have aimed at minimizing congestion and latency, for instance. 
In~\cite{dulinski2020dynamic}, the authors minimized financial expenses based on the $95^{th}$ percentile charging rule. 
Centralized and distributed solutions have been proposed to maximize network utility~\cite{allybokus2018multi, magnouche2021distributed}. 
While these solutions solve a load balancing problem for different  objectives (e.g., cost, fairness, latency), they do not aim at globally satisfying QoS requirements of applications.

\textbf{Adaptive QoS policies.} Active Queue Management (AQM) techniques have been proposed to meet delay and throughput requirements~\cite{adams2012active}. They use implicit (packet dropping) or explicit signals (ECN marking) to control sources sharing bottlenecks. However, they do not aim at tuning queuing parameters. Other works on Adaptive Weighted Fair Queuing (AWFQ) have proposed the dynamic adaptation of scheduling parameters. Some of them control only one device~\cite{FRANTTI200911390,sayenko2006comparison}. Others implemented a distributed protocol to trigger queue adjustments~\cite{HUSSAIN20031143} at neighborhood nodes. Most of existing methods do not optimize parameters globally. 
Also of interest, thanks to network calculus, worst case estimations can be derived for the end-to-end latency. 
Using the conservative latency-rate server model~\cite{latencyrate} for round-robin schedulers (e.g., WFQ, DRR), centralized network slicing algorithms~\cite{martin2021network} have been proposed to decide routing and bandwidth allocation for deterministic latency guarantees. However, they do not consider a hierarchical scheduler with low and high priority flows, as well as the overlay setting in which underlay resources are  hidden and uncertain. 

Adaptive mechanisms have been proposed in commercial SD-WAN solutions (see~\cite{adapiveQoS} for instance). Their main objective is to shape outgoing traffic over the WAN to avoid packet losses and congestion. The shaping rate is decreased when losses are detected inside the WAN and increased when losses are detected at egress queues. Some thresholds are used to stabilize the system. These solutions are reactive and do not try to estimate the available WAN bandwidth beforehand. They do not discriminate flow groups and blindly shape all outgoing traffic on each access routers. They also do not optimize shaping parameters globally as they uses local and greedy adaptations.

Our solution globally optimizes QoS policies using either a centralized or a distributed algorithm. It aims at protecting high priority flows and providing low priority
ones with a fair sharing of the remaining bandwidth in case
of heavy congestion. It also works in overlay scenarios where the available bandwidth is not known.

\textbf{Available bandwidth estimation.} 
Existing ABW methods~\cite{chaudhari2015survey} to estimate the available capacity on  overlay links can be categorized into: the Probe Gap Model (PGM) and the Probe Rate Model (PRM). PGM derives the available bandwidth by observing the intra-time gap of a pair of packets. IGI and Spruce are the most popular methods. These methods, however, require the knowledge of link capacities in the underlay.
PRM can address these drawbacks by observing the  self-induced congestion. Probing packets are injected into the networks and packet delay is measured at the receiver. If the probe rate is below the available bandwidth,  packets do not experience queuing delay. Otherwise, a high delay is detected. The most well-known PRM methods are PathLoad and PathChirp. Both of them face high estimation errors when the link utilization is high (i.e.  available bandwidth is low)~\cite{ali2007end}. Recently, Voyager~\cite{Voyager} has been introduced to improve accuracy. Using a decreasing-chirp-train method, it plots the best performance among all PRM methods. However, the error in case of high traffic (i.e. link utilization is over $90\%$) is still high. In all cases, these methods need to inject dedicated probes into the network and require an additional software at edge routers. As measurements about latency and packet loss are already widely available in commercial SD-WAN routers, we propose to exploit these data for ABW without injecting new probe traffic. 

Our solution goes beyond state-of-the-art as it defines a global framework to optimize queuing parameters on top of routing policies. It relies on the resolution of a global optimization problem to coordinate all edge devices in the context of overlay networks. We propose two solutions based on ABW to make informed decisions: a centralized one running inside the  controller and a distributed one that uses ABW as an implicit signal to coordinate edge devices.
Existing solutions do not consider a global optimization and do not rely on the estimation of the available bandwidth. 

\section{System Architecture}
\label{sec:SystemDescription}
We consider a semi-distributed architecture where edge devices are controlling traffic based on real-time measurements and policies are managed by a centralized controller. 

\subsection{SD-WAN use case}

\begin{figure}[t]
    \centering
    \includegraphics[width=0.95\columnwidth]{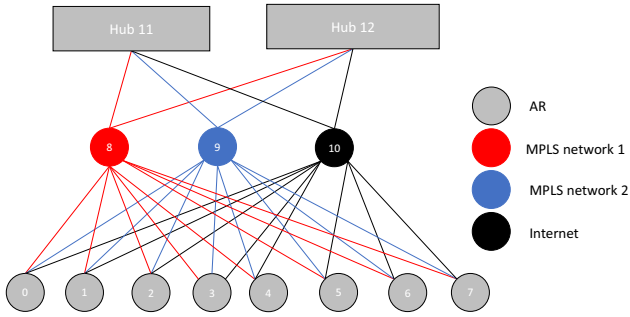}
    \caption{SDWAN Scenario with 2 hubs and 8 spokes.}
    \label{fig:sdwan}
\end{figure}

Fig.~\ref{fig:sdwan} presents a typical SD-WAN use case where $2$ data centers and $8$ branches (sites) of an enterprise are inter-connected by several networks (e.g. MPLS, Internet) controlled by third-party operators. A controller is placed at the headquarters site and branches are equipped with Access Routers (ARs). Flows from applications are aggregated in \emph{flow groups} that correspond to traffic classes with different SLA requirements. A typical traffic scenario includes Critical, VoIP, Office and Bulk flow groups, that respectively correspond to signaling, voice, business critical and non-critical applications.

\begin{figure}[b]
\centering
\includegraphics[width=1.0\columnwidth]{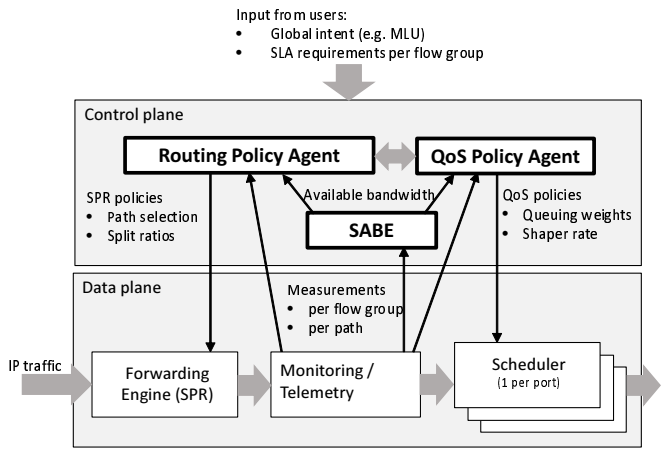}
\caption{Overall system architecture. 
}
\label{archi}
\end{figure}

Fig.~\ref{archi} depicts the overall system architecture. The architecture is split into two control entities operating at two
different time scales.
In a slow control loop, the control plane updates routing and QoS policies thanks to dedicated \emph{agents} and communicates them to the data plane. These agents are assisted by the SABE module for the estimation of the available bandwidth. In a fast control loop, the data plane takes tactical decisions to follow the evolution of traffic and network conditions. 
For each flow group, the policy  sent by the controller defines the routing logic (e.g., how to select outgoing paths), the QoS parameters (e.g., priorities, rate allocations) and security measures. The routing policy, for each flow group, consists in the set of overlay links that is allowed and its priority level, and 2) the QoS policy defines its minimum and maximum rate per outgoing overlay link.
Priorities are encoded using the differentiated services code point (DSCP) field of IP packets and, at each outgoing ports, a scheduler is used to discriminate high priority flows from low priority ones, and to enforce rate allocations.
Each AR router load balances traffic over the set of allowed overlay links according to actual network conditions (e.g. average delay, loss, jitter of overlay links). 

In our particular implementation, path selection is performed by AR routers  according to the SPR mechanism (see Sec.~\ref{sec:DeviceArchitecture}) and QoS policies are enforced by a hierarchical QoS architecture with a Class-Based Queuing (CBQ) scheduler per outgoing overlay link. Routing policies are optimized globally at the controller, while QoS policies can either be optimized at the controller or using a distributed algorithm (see Sec.~\ref{sec:Algorithms}).

\subsection{Smart Policy Routing}
\label{sec:DeviceArchitecture}

We briefly present how the Smart Policy Routing (SPR) mechanism works inside Huawei devices (see~\cite{huawei-SPR} for  details).
Access routers are configured with a policy for each flow group that contains the set of allowed overlay links and the SLA requirements that must be satisfied (i.e, acceptable latency, jitter and loss). 
The quality of overlay links is continuously monitored and evaluated. The set of overlay links that remain eligible with regard to the SLA requirements of each flow group is used to load balance traffic.
Traffic is distributed  proportionally to the nominal capacity of overlay links.

\subsection{Hierarchical QoS}
\label{sec:QueueArchitecture}

\begin{figure}[t]
    \centering
    \includegraphics[width=1.0\columnwidth]{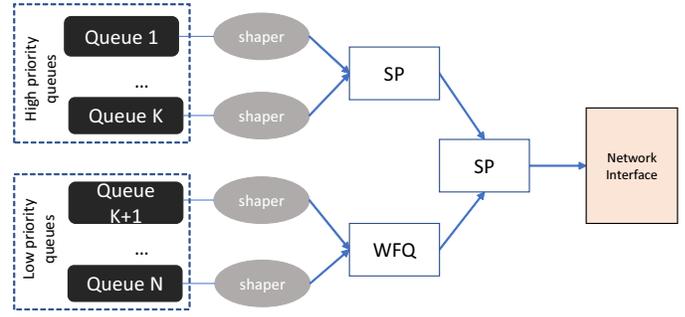}
    \caption{QoS architecture with a CBQ (Class-Based Queuing) scheduler per outgoing overlay link.}
    \label{fig:cbq}
\end{figure}

To further improve SLA satisfaction, beyond what load balancing can do, advanced QoS policies can be applied and adjusted dynamically.
The main target is first to 1) protect high priority flows and meet their requirements by controlling the bandwidth allocated to low priority ones, and then to 2) fairly share the remaining bandwidth among low priority flows. To achieve these goals, we propose to use 
a hierarchical QoS scheduler based on a CBQ scheduler per destination as depicted in Fig.~\ref{fig:cbq} presents. It handles high priority flows with Strict Priority Queuing (SP) and low priority flows with Weighted Fair Queuing (WFQ). The scheduler is deployed at each outgoing port of AR routers (at hubs and spokes).
The bandwidth allocation for the low priority flow groups is controlled thanks to shapers (i.e., maximum rate) and scheduling weights (i.e., minimum rate). For high priority flow groups, the shapers are not activated in order to guarantee the best SLA experience.

\subsection{Configuration of policies}
\label{}

In order to avoid congestion and satisfy SLA requirements, routing and QoS policies have to be carefully decided to mitigate interferences between flow groups. 
The policy optimization process takes as input measurements collected by monitoring: the average throughput of flow groups per min/hour/day/month, the average throughput and QoS of overlay links, the estimation of the available bandwidth with SABE, and  the SLA satisfaction of each flow group over each overlay link. It also considers the nominal capacity of overlay links, the global intents to optimize and the SLA requirements of each flow group. The control plane  also uses priorities of flow groups and their preferences for each transport network.

As presented in a previous work~\cite{quangintent}, the controller optimizes SPR policies (load balancing) to optimize SLA satisfaction and global intents (e.g., minimizing the Maximum Link Utilization - MLU). In order to take non-myopic decisions, this process can also take as input traffic and SLA predictions. On top of that, we propose in this paper to adjust QoS policies: we present a centralized solution that can operate inside the controller and a distributed solution where AR routers collaborate through ABW measurements.

\section{Policy Optimization Model}
\label{sec:ProblemFormulation}

We present policy optimization models for both SPR and QoS policies to enhance SLA satisfaction for all flow groups. 

\textbf{Problem formulations.}
We consider a set of overlay links $E$ of capacity $C_e$. For each flow group $k\in K$, the measured traffic demand is denoted $b^k$.  The delay and packet loss requirements of the flow group $k$ are denoted $D_k$ and $L_k$, respectively. $d_e^k$ and  $l_e^k$ are delay and packet loss  functions for flow group $k$ on overlay link $e$. In this paper, these values are determined by actual measurements, but the function can return predictions depending on  split ratios $x_e^k$, i.e., the fraction of traffic of the flow group $k$ that is sent over the link $e$. 

\textbf{1) SPR policy optimization problem.} The following model optimizes a global intent (i.e. minimize the maximum link utilization) while meeting SLAs  (see~\cite{quangintent} for model details). It decides split ratios given by $x_e^k$ variables.

\begin{alignat}{3}
&\min \quad & \alpha_s \sum_{k\in K} \left(u_k+v_k\right) + \beta_s LU +\nonumber \\ 
& \quad & + \gamma_s \sum_{e\in E} \sum_{k\in K} \left(d_e^k(x) + l_e^k(x) \right)
\label{objSPR}
\end{alignat}
\begin{alignat}{3}
& %\sum_{i \in I_e} 
\sum_{k\in K} b_k x_e^k \leq LU.C_e &  &\quad \forall e\in E, \label{capacity_constraints}\\
  & d_e^k(x) \leq D_k + u_k  & &\quad \forall k\in K,\forall e\in E,   \label{delay_constraints}\\
  & l_e^k(x) \leq L_k +v_k & &\quad \forall k\in K,\forall e\in E,   \label{loss_constraints}\\
  & \sum_{e\in E} x_e^k =1& &\quad \forall k\in K   \label{convexity_constraints}
\end{alignat}

$\alpha_s$, $\beta_s$ and $\gamma_s$ are weights corresponding to the $3$ parts of the objective function: the minimization of SLA violations defined by $u_k$ and $v_k$ for the delay and packet loss (most important), the minimization of the maximum Link Utilization (LU)  and the minimization of the delay and packet loss to improve SLAs beyond requirements (least important). 

\textbf{2) QoS policy optimization problem.} Once SPR policies have been optimized through a \textit{slow control loop}, the traffic for flow group $k$ over overlay link $e$ is expected to be $d^k_e = b_k x_e^k$. The set of all allowed overlay links is given by $E_k\subseteq E$ (i.e., overlay links with non zero $x_e^k$). To further optimize bandwidth sharing at each port, QoS policies are optimized over a \textit{faster control loop} thanks to the following model that decides queuing parameters for low priority flow groups.

\begin{alignat}{3}
\min \quad & \sum_{e,k} \left(\alpha_q h^k_e - \beta_q d^k_e \log{z^k_e} +\gamma_q \left(y_e^k+v_e^k \right)z_e^k  \right)  \label{obj_QoS}\\
\textrm{s.t.} \quad & \sum_{k,e' \in \mathcal{F}(e) \cup {e} } z^k_{e'} \leq C_e ,  \quad \forall e\in E, \label{capacity_constraints}\\
  & d_e^k(z) \leq D_k + y_e^k \quad \forall k\in K,\forall e\in E_k,   \label{delay_constraints}\\
    & l_e^k(z) \leq L_k + v_e^k \quad \forall k\in K,\forall e\in E_k,   \label{delay_constraints}\\
  & z^k_e+h^k_e \geq d^k_e,\quad \forall k\in K,\forall e\in E   \label{demand_constraints}
\end{alignat}

Similarly to SPR, $\alpha_q$, $\beta_q$ and $\gamma_q$ are weights corresponding respectively to the $3$ parts of the objective function: the minimization of rejected traffic calculated with variables $h_e^k$ that measure the gap between the traffic demand and the allocated rate (most important), the fair allocation of bandwidth among flow groups using standard proportional fairness~\cite{kelly1998rate}, and the minimization of SLA violations defined by $y_e^k$ and $v_e^k$ for the delay and the packet loss (least important). The output is the rate allocation for each flow group over each overlay link, $z^k_e$, which is used to tune queuing weights and shaping. The normalized values of $z$ define WFQ weights and original $z$ values defines shaping rates.

\section{SABE: Silent Available Bandwidth Estimation}
\label{sec:ABW}

Measurements about latency and packet loss for overlay links are already widely available in commercial SD-WAN routers, e.g. using Network Quality Analysis (NQA)~\cite{huawei-NQA} in Huawei devices. Therefore, we introduce SABE (Silent Available
Bandwidth Estimation), a new ABW technique that only exploits
available measurements.
This method leverages on a simple M/M/1/K queuing model that considers each overlay link as a queuing system with a single server (the WAN network). It assumes that incoming traffic follows a Poisson process and that the  service time follows an exponential distribution. The capacity of the queue is of $K$ packets.

\subsection{M/M/1/K queuing model}

From \cite{sztrik2012basic}, the packet loss rate in M/M/1/K model is determined as follows:
\begin{equation}
P=\begin{cases}
     \frac{\left(1-\rho\right)\rho^{K}}{1-\rho^{K+1}} & \rho \neq 1\\
      \frac{1}{K+1} & \rho=1 \textrm{ otherwise}
    \end{cases},
\label{eq:pkloss}
\end{equation}
where $\rho$ is the mean load of the server. $\rho$ can be derived from the mean service rate $\mu$ (packet/s) and the mean arrival rate $\lambda$ (packet/s) with $\rho=\frac{\lambda}{\mu}$. 

The mean delay for M/M/1/K model is computed as follows
\begin{equation}
    D=\frac{L}{\lambda_e},
\label{eq:delay}
\end{equation}

where $\lambda_e$ is the effective server utilization and derived from the arrival rate and the probability of packet loss from Eq.~\ref{eq:pkloss} as $\lambda_e = \lambda (1-P)$.  $L$ is the average number of packets in the system and can be determined from the server load as follows 
\begin{equation}
    L=\begin{cases}
    \frac{\rho}{1-\rho}-\frac{(K+1)\rho^{K+1}}{1-\rho^{K+1}}& \rho \neq 1 \\
    \frac{K}{2} & \rho =1
    \end{cases}
\end{equation}

\subsection{Available bandwidth estimation}

\begin{figure}
    \centering
    \includegraphics[width=0.9\columnwidth]{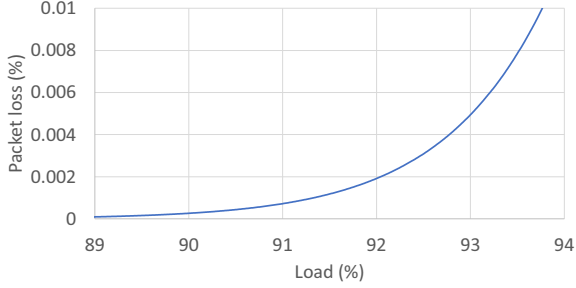}
    \caption{Packet loss vs load in an M/M/1/K queue with K=100. 
    }
    \label{fig:pkloss}
\end{figure}

The total load $\rho_e$ of an overlay link $e$ is the sum of  1) the traffic $X_e$ that AR routers inject, which is known by the SD-WAN system, and 2) the cross-traffic $XT_e$, that is out of control. Using the queuing model presented above, we can estimate the overlay link load $\rho_e$ from already available delay and loss measurements on WAN links. This method is \emph{silent} as it only exploits available data.

Indeed, it is possible to estimate the overlay link load $\rho_e$ (i.e., arrival rate) from Eq.~\ref{eq:pkloss} and~\ref{eq:delay}. These two equations are functions of $\rho=\frac{\lambda}{\mu}$. The service rate $\mu$ can be estimated considering a typical packet size, of $ps$ bits, and the nominal link capacity, $C_e$ in Mbps, i.e. $\mu=\frac{ps}{C_e}$. Finally,  Eq.~\ref{eq:pkloss} and~\ref{eq:delay} can then be inverted to recover $\rho_e$ from the measured delay $D_e$ and packet loss rate $L_e$. This operation is actually non-trivial. To solve this issue, we discretize the load and look for the $\lambda$ corresponding to delay or packet loss as close as possible to measurements. When no WAN loss is observed, i.e. $L_e = 0$, Eq.~\ref{eq:delay} is used. In the other case, we resort to Eq.~\ref{eq:pkloss}. The procedure repeats periodically.

Once the overlay link load $\rho_e$ is estimated, the available bandwidth $\widetilde{C}_e$ can be finally estimated as follows:

\begin{equation}
    \widetilde{C}_e = \theta_e. C_e - XT_e \textrm{, where } XT_e = \rho_e - X_e
\label{eq:SABE}
\end{equation}

where $\theta_e$ is a down-scaling parameter that we  use to ensure safety, i.e. to avoid packet losses inside the WAN. In fact, from Eq.~\ref{eq:pkloss}, as depicted in Fig.~\ref{fig:pkloss}, we can analyze the relationship between the load and the packet loss. We considered $K=100$ as a realistic lower bound for the size of queues in today's routers. From the figure, we can observe  that we can limit the maximum link utilization below a given $\theta_e$ to keep packet loss under a given threshold. For instance, the value of $\theta_e$ should be $0.914$ to guarantee a packet loss lower than $0.001\%$.

\section{Local Search Algorithm}
\label{sec:Algorithms}
In this section, we propose a local search algorithm to calculate a good QoS policy update at each step (each time the algorithm is called). The goal of this algorithm is to evaluate all possible modifications of the shaper rates of low priority flow groups on overlay links (i.e., QoS policies) and select the best one in an iterative manner. The algorithm continues until the solution cannot  improve. We present two versions: a centralized one running inside
the controller and a distributed one that uses ABW as an implicit signal to coordinate routers. 

\textbf{Centralized version}. The centralized algorithm is described in Alg.~\ref{alg:cenLS}. To determine the rate allocation of each flow group $k$, the algorithm exploits measurements on each overlay link for each flow group about delay, packet loss and throughput. It also leverages on the estimation of the available bandwidth derived from SABE as discussed in Sec.~\ref{sec:ABW}. 
The rate of high priority flow groups is allocated first. If capacity remains, it is shared among low priority flow groups using objective function of Eq.~\ref{obj_QoS}. The algorithm finds the pair of flow group and overlay link, i.e. ($k_0, e_0$), with the greatest contribution to the objective function ($Obj_e^k$) and attempts to decrease its contribution by increasing its allocated rate of an amount $\delta$. The algorithm keeps increasing the rate for all low priority flow groups until the objective cannot be improved. 

\textbf{Distributed version}. 
In the distributed algorithm, each AR router estimates the available bandwidth for each WAN link using SABE. Then, it executes Alg.~\ref{alg:cenLS} with  local inputs only (i.e. flow groups and overlay links starting / ending at the AR). \jeremie{ABW is used as an implicit signal about how traffic other AR routers are sending.}

\begin{algorithm}[t]
\small
	\SetAlgoLined
	\KwResult{Rate allocations $z$ of flow groups on overlay links}
	$RemC_{e}$: remaining capacity of  link $e$, initialized as $\widetilde{C}_{e}$\; 
	$Obj_e^k = \alpha_q h^k_e - \beta_q d^k_e \log{z^k_e} +\gamma_q \left(y_e^k+v_e^k \right)z_e^k $, the contribution of flow group $k$ on link $e$ to the objective;
	
	$\tilde{\mathcal{E}} =\left\{e : RemC_{e'} \geq \delta, \forall e' \in \mathcal{F}(e)\cup \{e\}\right\}$, set of links where the rate can be increased by $\delta$ for a flow group\;
 \tcc{First, high priority flow groups }
	\For{each overlay link $e$}{
	    $\tilde{\mathcal{E}} \leftarrow e$\;
	    \For{each high priority flow group $k$}{
	    \tcc{rate allocation depending on the demand or capacity}
	        $z_e^k = \min \left(d_e^k, \tilde{c}_e\frac{d_e^k}{\sum_{k,e' \in \mathcal{F}(e) \cup \{e\} } d_{e'}^{k}}\right)$\;
	        \tcc{Update remaining capacities}
	        \For {$e' \in \mathcal{F}(e) \cup \{e\}$}{
	            $RemC_{e'} = RemC_{e'} - z_{e}^{k}$\;
	        }
	    }
	}
	\tcc{Second, low priority flow groups}
	\While{ \text{true} }{
		\tcc{Update $\tilde{\mathcal{E}}$}
		\For {each overlay link $e$}{
		    \If {$\exists e' \in \mathcal{F}(e) \cup {e} : RemC_{e'} < \delta $}{
		    $\tilde{\mathcal{E}} \smallsetminus \left\{e\right\}$\;
		    }
		}
        \quang{
        \If {$\tilde{\mathcal{E}}$ is empty}{
            \tcc{No more spare bandwidth}
            break\;
        }
        }
		\tcc{Attempt to increase rates}
		$\left(k_0,e_0\right) \leftarrow \max_{k,e \in \tilde{\mathcal{E}}} Obj_e^k$\;		
         $z_{e_0}^{k_0} \leftarrow z_{e_0}^{k_0}+ \delta $ \;
        \For {$e' \in \mathcal{F}(e_{0}) \cup {e_{0}}$}{
            $RemC_{e'} = RemC_{e'} - \delta$\;
        }
	}
	\caption{Local search algorithm - Centralized 
	}	\label{alg:cenLS}
\end{algorithm}

\section{Simulation Results}
 \label{sec:NumericalResults}  
 We now evaluate our SD-WAN architecture using NS3~\cite{ns3} with OpenFlow 1.3~\cite{Chaves:OF13} to emulate the control plane logic, select routing paths and adjust queuing parameters. In the data plane of OpenFlow switches, a CBQ  scheduler per destination is implemented with shapers for low priority flow groups. As depicted in Fig.~\ref{fig:sdwan}, the simulation is composed of two data-centers connected to $8$ sites with 2 MPLS and 1 Internet providers. There are $16$ origin-destination pairs, all  from data-centers to spokes. MPLS lines have propagation delays of $10$ms and $20$ms and capacities of $5$ Mbps and $10$ Mbps, respectively. Internet lines have propagation delays of $30$ms and capacities of $25$ Mbps, respectively. Note that capacity and traffic have 
been down scaled compared to reality so that execution time
is  acceptable.
We considered four types of applications: Critic, VoIP, Office, and Bulk. SLA requirements of these flow groups are shown in Table~\ref{tab:qosreq}. The transport layer is TCP. Microflow inter-arrival time varies to generate diurnal traffic patterns. \jeremie{We simulated a traffic scenario over 1200 seconds.}

The controller computes the global SPR policy every $100$s, while the global QoS policy is computed every $10$s. In the distributed  scenario, QoS policies are computed by each router, every $10$s also. Note that SPR and QoS policies can also be computed in an event-triggered manner (e.g. when the congestion is detected). However, in the context of this paper, we consider only time-triggered SPR and QoS policy optimization.
 \begin{figure}
     \centering
    \begin{subfigure}[b]{0.5\textwidth}
     \centering
     \includegraphics[width=\textwidth]{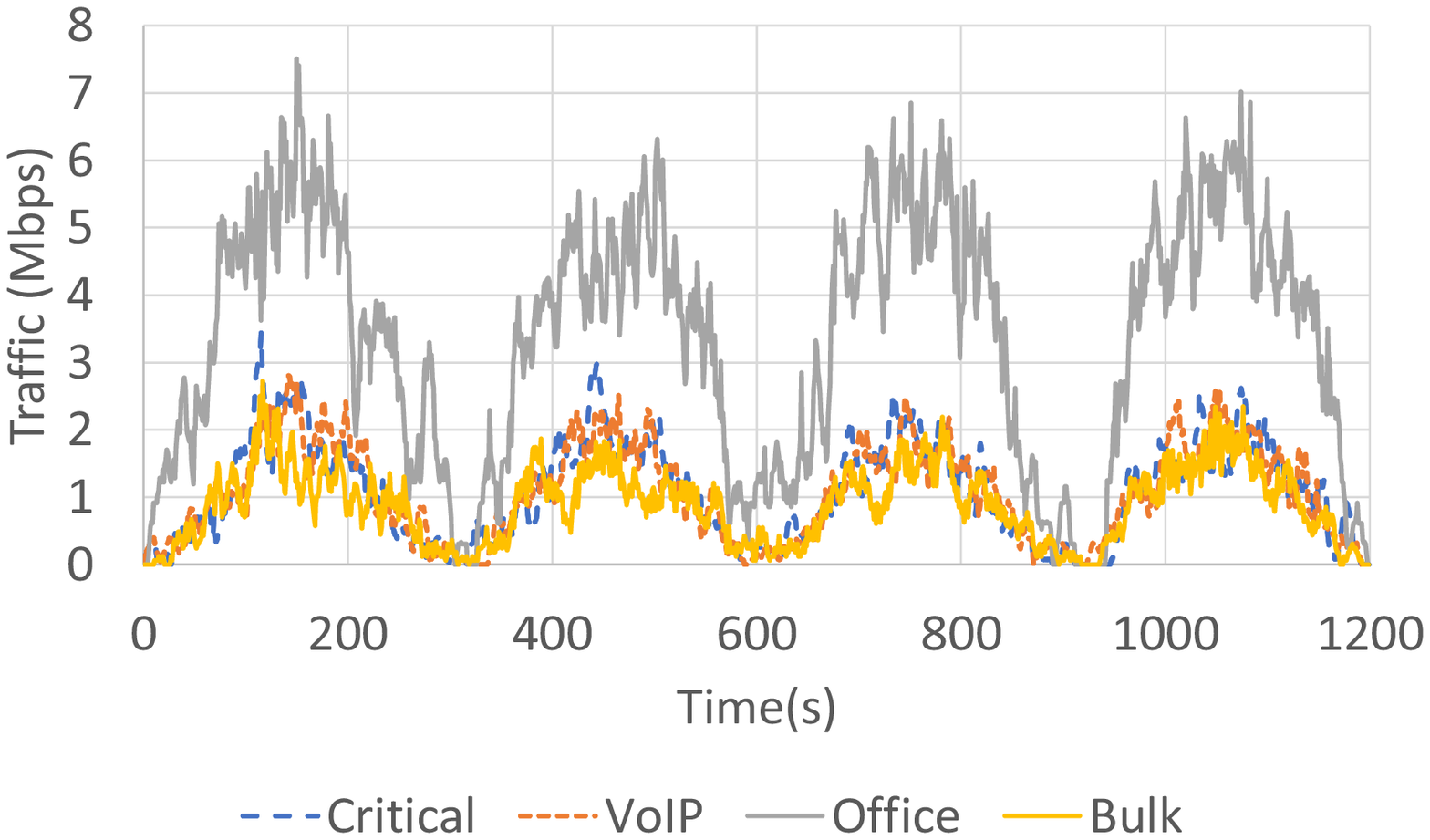}
     \caption{Traffic in each flow groups for each OD.}
     \label{fig:ODTraffic}
    \end{subfigure}
    \begin{subfigure}[b]{0.5\textwidth}
     \centering
     \includegraphics[width=\textwidth]{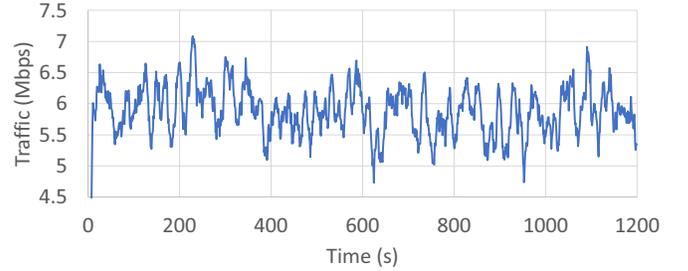}
     \caption{Cross-traffic on the Internet}
     \label{fig:ODTraffic}
    \end{subfigure}
         \caption{Traffic per OD and cross-traffic on the Internet.}
\end{figure}

 \begin{table}[]
     \centering
     \begin{tabular}{|l|c|c|c|c|}
        \hline
        Flow group & Critical & VoIP & Office & Bulk  \\
        \hline
        Delay requirement (ms)  & 40 & 60 & 250 & 800 \\
        \hline
        Packet loss requirement ($\%$) & 0 & 2 & 5 & 10 \\
        \hline
     \end{tabular}
     \caption{SLA requirements (maximum values).}
     \label{tab:qosreq}
 \end{table}
 
 \subsection{Available bandwidth estimation accuracy}
\begin{figure}
     \centering
     \includegraphics[width=1.0\columnwidth]{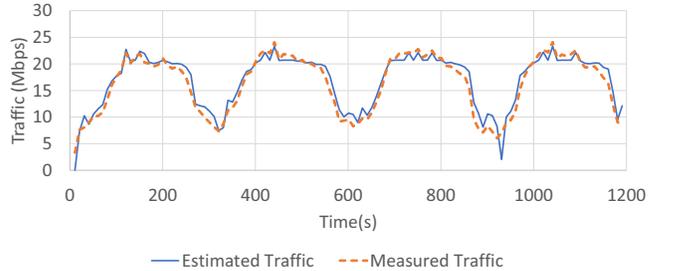}
     \caption{Estimated traffic vs Measured Traffic with SABE.}
     \label{fig:ABW}
 \end{figure}

 The monitoring module measures the quality of the overlay link (i.e. delay, jitter, packet loss) every $10$ seconds. The available bandwidth on overlay links can be estimated with SABE based on the measurements. Fig.~\ref{fig:ABW} presents the estimated traffic with SABE and the total traffic measured on the overlay link (i.e. the ground truth). The average relative error is $9.33\%$ \jeremie{and the maximum relative error is up to $20\%$. While it deserves a deeper evaluation in real conditions, the relative error obtained in our simulation scenario is comparable with state of the art ABW methods~\cite{Voyager} based on active probing. However, SABE fully relies on already available measurements between access routers for packet loss and delay.}
 
 \subsection{Benchmark solutions}
We consider multiple solutions for the optimization of SPR and QoS policies. We tested two SPR policies: (i) \allTns~ and (ii) \mlu. The All Transport Network (\allTns)~policy allows ARs to use all available  networks (e.g. MPLS and Internet), while the \mlu~policy optimizes SPR policies to minimize the MLU (as presented in~\cite{quangintent}). For  QoS policies, we consider several solutions: (i) Fix Weights (\texttt{FW}) and (ii) the optimization of QoS parameters solving the optimization problem (Sec.~\ref{sec:ProblemFormulation}) with either a centralized Local Search algorithm (\texttt{LS}), a distributed one (\texttt{Dist}) or a commercial solver (\texttt{OPT}).  The exact optimal QoS policy \texttt{OPT} is derived using a non-linear solver like SCIP~\cite{scip}. Note that in practice, this solution takes too much time, especially in large scale scenarios. 

\texttt{FW} determines the weights for WFQ based on the peaks of traffic. From \cite{latencyrate}, the delay of queue $k$ (dedicated to flow group $k$) with weight $w_k$ is bounded as follows,
 \begin{equation}
     d_e^k \geq \frac{z_e^k}{w_k C_e},
 \end{equation}
 where $z_e^k$ is the traffic of flow group $k$ on overlay link $e$. As $d_e^k$ should be less than or equal to the delay requirement $D_k$, we have $D_k \geq \frac{z_e^k}{w_k C_e}$. Therefore, $w_k \geq \frac{z_e^k}{D_k C_e}$. Higher weights give more bandwidth to the flow group, thus the delay can go beyond the requirement. It can, however, impact negatively other flow groups. Consequently, we select $w_k = \frac{z_e^k}{D_k C_e}$.

\begin{table*}[!htbp]
\centering
\begin{tabular}{|P{1cm}||P{1.25cm}|P{1.7cm}|P{1.7cm}|P{1.7cm}|P{1.25cm}|P{1.8cm}|P{1.8cm}|P{1.7cm}|}
 \hline
  \diagbox[width=\dimexpr \textwidth/18+2\tabcolsep\relax, height=0.7cm]{App}{SLA} &\multicolumn{4}{c|}{Packet loss satisfaction rate (\%) (without/with SABE)}&\multicolumn{4}{c|}{Delay satisfaction rate (\%) (without/with SABE)}\\
  \hline
 &\allTNsFW&\allTNsScip&\allTNsLS&\Dist&\allTNsFW&\allTNsScip&\allTNsLS&\Dist \\
 \hline
Critical & 54.96  & 60.80 / 85.24  & 59.40 / 84.15  &55.46 / 68.56 & 55.63 & 61.22 / 86.07  & 84.74 / 84.15& 56.71 / 70.40\\
\hline
VoIP & 60.72 & 75.48 / 97.41  & 75.58 / 98.33 & 62.30 / 80.57 & 77.65 & 90.66 / 99.67& 98.58 / 99.92 & 81.98 / 95.83\\
\hline
Office & 96.16 & 97.16 / 99.58 & 96.33 / 99.42 & 97.41 / 91.90 & 100.00 & 84.74 / 98.58& 98.42 / 98.50& 72.23 / 75.20\\
\hline
Bulk & 96.50 & 95.75 / 99.25 & 98.33 / 98.25 & 98.08 / 96.08 & 100.00 & 99.75 / 98.50& 98.50 / 98.25& 82.05 / 86.20\\
\hline
 \diagbox[width=\dimexpr \textwidth/18+2\tabcolsep\relax, height=0.7cm]{App}{SLA}&\multicolumn{4}{c|}{Average APP Packet loss  (\%) (without/with SABE)} &\multicolumn{4}{c|}{Average Delay (s) (without/with SABE)}\\
 \hline
Critical & 2.72 & 1.44 / 0.19   &  1.52 / 0.27   &  2.42 / 1.15 & 0.042 & 0.038 / 0.03 & 0.03 / 0.03 &0.042 / 0.035 \\
\hline
VoIP & 2.69 & 1.30 / 0.16  & 1.42 / 0.16 & 2.37 / 1.01 & 0.042 & 0.037 / 0.03& 0.03 / 0.03& 0.041 / 0.033 \\
\hline
Office & 2.08 & 1.99 / 1.86 & 1.30 / 1.90 & 2.12 / 2.64 & 0.043 & 0.086 / 0.120& 0.12 / 0.13 & 0.230 / 0.410\\
\hline
Bulk & 2.49 & 1.83 / 2.53 & 1.99 / 2.69 &  2.06 / 2.62 & 0.044 & 0.240 / 0.380& 0.39 / 0.41 & 0.450 / 0.660\\
\hline
\end{tabular}
\caption{SLA performances for All-TNs SPR policy and the different QoS policies.}
\label{tab:pkloss-alltn}
\end{table*}

\subsection{Performance evaluation}
%comments on AllTNs + QoS
Table~\ref{tab:pkloss-alltn} shows the SLA performance when the SPR policy is \allTns~\jeremie{(i.e., when SPR policy are not optimized). The table presents for all flow groups (i.e., applications), the average performance and SLA satisfaction rate for the end-to-end delay and packet loss. In each cell, performance are presented with and without the use of SABE to tune capacity constraints related to overlay links.} As we can see, 
\allTNsFW~plots the worst performance for high priority flow groups. The high traffic load of Office (see Fig.~\ref{fig:ODTraffic}) yields to high queuing weights, therefore, \allTNsFW~gives a high priority to low priority traffic and this leads to congestion  which degrades the SLA satisfaction rate of high priority traffic. For \allTNsFW, the SLA satisfaction rates of low priority flow groups are as high as $90\%$ to $100\%$, \jeremie{while it is around $55\%$ for high priority flow groups. At the other extreme,} \allTNsScip~is able to protect SLA for high priority flow group much better \quang{as it \jeremie{dynamically tune} the weights (i.e., min rate) and the shaper rates (i.e., max rate) of each queue based on SLA requirements\jeremie{, priorities and the actual traffic demand}. For high priority flow groups (Cirtical and VoIP), the gain in terms of packet loss satisfaction rate is about $5\%$ to $15\%$ and it is extended to $30\%$ when using SABE thanks to awareness of background traffic. \jeremie{Similar gains can be observed for the end-to-end delay.} However, the optimal solution of the non-linear optimization problem in \allTNsScip~ is intractable and the local search algorithm proposed in Sec.~\ref{sec:Algorithms} is a practical solution. \allTNsLS~offers similar performance in terms of packet loss satisfaction rate to \allTNsScip. There is a negligible loss in performance when using local search, e.g. $84.15\%$ for \allTNsLS~with SABE versus $85.24\%$ for \allTNsScip~with SABE. In \Dist~without SABE, each AR determines the rate allocation based on the local information it has. Due to a wrong estimation of what is happening on the WAN, the SLA satisfaction rates are low and similar to \allTNsFW. When ABW is used and integrated in \Dist, AR can have a better estimation of background traffic in WAN \jeremie{and of the traffic that other AR are sending. Thanks to this, we can observe an improvement} of the packet loss satisfaction rate by $13\%$ in the critical flow group.}

\begin{table*}[!htbp]
\centering
\begin{tabular}{|P{1cm}||P{1.25cm}|P{1.7cm}|P{1.7cm}|P{1.7cm}|P{1.25cm}|P{1.8cm}|P{1.8cm}|P{1.7cm}|}
 \hline
  \diagbox[width=\dimexpr \textwidth/18+2\tabcolsep\relax, height=0.7cm]{App}{SLA} & \multicolumn{4}{c|}{Packet loss satisfaction rate (\%) (without/with SABE)} &\multicolumn{4}{c|}{Delay satisfaction rate (\%) (without/with SABE)}\\
 \hline
 &\mluFW& \mluScip & \mluLs&\Dist &\mluFW& \mluScip & \mluLs&\Dist \\
 \hline
Critical & 54.63  & 64.39 / 97.75  & 63.75 / 97.66  & 53.29 / 89.41 & 76.23 & 79.32 / 97.75 & 79.9 / 97.83 & 76.81 / 91.25 \\
\hline
VoIP & 60.55 & 99.08 / 100.00  & 100.00 / 99.75 & 61.55 / 97.91 & 96.25 & 99.92 / 100.00 & 100.00 / 100.00 & 95.75 / 100.00\\
\hline
Office & 98.58 & 100.00 / 99.25 & 98.65 / 99.42 & 98.41 / 98.83 & 100.00 & 100.00 / 100.00 & 100.00 / 100.00 & 76.65 / 87.72 \\
\hline
Bulk & 100 & 89.16 / 90.33 & 89.58 / 95.33 & 100.00 / 99.67 & 100.00 &100.00 / 100.00 & 99.92 / 83.14& 82.60 / 91.30\\
\hline
   \diagbox[width=\dimexpr \textwidth/18+2\tabcolsep\relax, height=0.7cm]{App}{SLA}&\multicolumn{4}{c|}{Average APP Packet loss (\%) (without/with SABE)} &\multicolumn{4}{c|}{Average Delay (s) (without/with SABE)} \\
 \hline
Critical & 1.60  & 0.26 / 0.010   & 0.013 / 0.012 & 1.53 / 0.144 & 0.034 & 0.032 / 0.029 & 0.030 / 0.029 & 0.034 / 0.029\\
\hline
VoIP & 1.62 & 0.230 / 0.037  & 0.230 / 0.024  &  1.54 / 0.117 & 0.034 & 0.032 / 0.029 & 0.030 / 0.029 & 0.034 / 0.027 \\
\hline
Office & 2.31 & 1.87 / 2.580 & 3.450 / 2.560 & 2.57 / 1.600 & 0.036 & 0.069 / 0.134 & 0.115 / 0.149 & 0.194 / 0.185\\
\hline
Bulk & 0.12 & 2.28 / 3.170  &  2.100 / 3.100  & 0.70 / 2.190 & 0.032 & 0.233 / 0.326 & 0.382 / 0.530 & 0.410 / 0.480 \\
\hline
\end{tabular}
\caption{SLA performances for \mlu~SPR policy and all the different policies.}
\label{tab:pkloss-mlu}
\end{table*}

%comments on joint SPR+QoS
The performance when SPR policies are optimized with \mlu~is shown in Table~\ref{tab:pkloss-mlu}. \jeremie{In this case, the local search algorithm presented in ~\cite{quangintent} is used to periodically optimize policies to minimize the MLU. The goal is to further improve performance with a better load balancing. }
When the \mlu~policy is applied on top of all QoS policies and SABE, \quang{the SLA satisfaction of critical flow groups is improved by $12\%$ compare to \allTNsScip~and by $40\%$ compared to static policies in \allTNsFW. This improvement comes from a better global load balancing of traffic over the multiple transport networks. QoS policies are designed to avoid the congestion inside the WAN by limiting traffic but it cannot shift traffic from one transport network to another. 
Consequently, to obtain the best SLA satisfaction, the combination of \mlu~SPR, QoS optimization, and ABW is necessary. By deploying them together, both delay and packet loss requirements can be guaranteed (satisfaction rate $>95\%$)}.

\quang{In all the above results, we have seen that SABE plays an important role in providing an accurate estimation of the available bandwidth, therefore improving SLA satisfaction. In the distributed solution,  SABE} helps ARs to estimate cross-traffic and the traffic injected by other AR devices. The distributed algorithm without SABE has much lower SLA performance in all configurations. \quang{However, when the distributed solutions is used in combination with  SABE, the SLA satisfaction rate is comparable with the centralized solution (the SLA satisfaction rate of the distributed solution with SABE is up to $20\%$ lower than for the centralized solution with SABE)}.
\section{Conclusion}
\label{sec:conclusion}
We proposed centralized and distributed QoS policy optimization system for SD-WAN that is able to protect the quality of high priority flow groups while avoiding the bandwidth starvation for low priority flow groups. The proposed QoS policy optimization is built on top of existing SPR policy optimization~\cite{quangintent} for load balancing and plays a complementary role in optimizing SLA performance. The proposed ABW mechanism leverages monitoring data that are already available to infer the available bandwidth on WAN links, therefore it avoids adding complexity to existing systems. We demonstrated through simulations that the combination of load balancing, QoS optimization, and ABW can significantly help to guarantee SLAs in both centralized and semi-distributed settings \jeremie{(i.e., where SPR policies are optimized centrally and QoS policies are optimized distributively)}

\jeremie{Future work along these lines include a more extensive evaluation of the silent ABW technique we proposed (i.e., SABE). We also would like to develop advanced distributed solutions that may leverage on some limited exchanges between AR routers to further improve performance and the reactivity to abnormal events.}
 
\bibliographystyle{IEEEtran}
\bibliography{ref.bib}
\end{document}